\documentclass[pdflatex,sn-mathphys-num]{sn-jnl}
\usepackage{graphicx}%
\usepackage{multirow}%
\usepackage{amsmath,amssymb,amsfonts}%
\usepackage{amsthm}%
\usepackage{mathrsfs}%
\usepackage[title]{appendix}%
\usepackage{xcolor}%
\usepackage{textcomp}%
\usepackage{manyfoot}%
\usepackage{booktabs}%
\usepackage{url}
\usepackage{algorithm}%
\usepackage{algorithmicx}%
\usepackage{algpseudocode}%
\usepackage{listings}%
\usepackage{comment}
\usepackage{anyfontsize}
\usepackage[T1]{fontenc}
\usepackage{verbatim}

\theoremstyle{thmstyleone}%
\theoremstyle{thmstyletwo}%

\theoremstyle{thmstylethree}%

\begin{document}

\title[Article Title]{Reconfigurable field-free spin Hall nano-oscillators enabled by crystallographic anisotropy in epitaxial Co/Pt}


\author*[1]{\fnm{Jong-Guk} \sur{Choi}}\email{jongguk.choi@physics.gu.se}
\equalcont{These authors contributed equally to this work.}

\author[1,2]{\fnm{Avinash Kumar} \sur{Chaurasiya}}
\equalcont{These authors contributed equally to this work.}

\author[3]{\fnm{Jaimin} \sur{Kang}}
\equalcont{These authors contributed equally to this work.}

\author[1]{\fnm{Venkatesh} \sur{Vadde}}
\author[4]{\fnm{Peter G.} \sur{Lim}}
\author[1]{\fnm{Roman} \sur{Khymyn}}
\author[1,5,6]{\fnm{Ahmad A.} \sur{Awad}}
\author[1,5,6,7]{\fnm{Akash} \sur{Kumar}}
\author[3,4,8,9]{\fnm{Mark C.} \sur{Hersam}}
\author[4,8]{\fnm{Vinayak P.} \sur{Dravid}}
\author*[3,4]{\fnm{Pedram Khalili} \sur{Amiri}}\email{pedram@northwestern.edu}
\author*[1,5,6]{\fnm{Johan} \sur{Åkerman}}\email{johan.akerman@physics.gu.se}

\affil[1]{\orgdiv{Department of Physics}, \orgname{University of Gothenburg}, \orgaddress{\street{Fysikgränd 3}, \city{ Gothenburg}, \postcode{412 96}, \country{Sweden}}}

\affil[2]{\orgdiv{Department of Physics}, \orgname{Indian Institute of Science Education and Research Bhopal}, \orgaddress{\city{ Bhopal}, \postcode{462 066}, \country{India}}}

\affil[3]{\orgdiv{Department of Electrical and Computer Engineering}, \orgname{Northwestern University}, \orgaddress{\street{Evanston}, \city{ Illinois}, \postcode{60208}, \country{USA}}}

\affil[4]{\orgdiv{Applied Physics Program}, \orgname{Northwestern University}, \orgaddress{\street{Evanston}, \city{ Illinois}, \postcode{60208}, \country{USA}}}

\affil[5]{\orgdiv{Research Institute of Electrical Communication (RIEC)}, \orgname{Tohoku University}, \orgaddress{\street{2-1-1 Katahira, Aoba-ku}, \city{Sendai}, \postcode{980-8577}, \country{Japan}}}

\affil[6]{\orgdiv{Center for Science and Innovation in Spintronics (CSIS)}, \orgname{Tohoku University}, \orgaddress{\street{2-1-1 Katahira, Aoba-ku}, \city{Sendai}, \postcode{980-8577}, \country{Japan}}}

\affil[7]{\orgdiv{Centre for Interdisciplinary and Convergent Technologies}, \orgname{Indian Institute of Technology Kharagpur}, \orgaddress{\city{Kharagpur}, \postcode{721302}, \country{India}}}

\affil[8]{\orgdiv{Department of Materials Science and Engineering}, \orgname{Northwestern University}, \orgaddress{\street{Evanston}, \city{ Illinois}, \postcode{60208}, \country{USA}}}

\affil[9]{\orgdiv{Department of Chemistry}, \orgname{Northwestern University}, \orgaddress{\street{Evanston}, \city{ Illinois}, \postcode{60208}, \country{USA}}}

\abstract{Spin Hall nano-oscillators (SHNOs) are nanoscale microwave sources for wireless communication, neuromorphic computing and oscillator-based Ising machines, but conventional devices require a global magnetic bias. Here we replace this bias through crystallographic anisotropy in epitaxial Co/Pt. Growth of hcp Co with its \(c\)-axis in the film plane produces an anisotropy field of about \(0.36~\mathrm{T}\) and enables field-free auto-oscillations
above \(10~\mathrm{GHz}\) in nanoconstriction SHNOs. The active current polarity is selected by the remanent magnetization, providing nonvolatile reconfiguration of the oscillation state. Micro-focused Brillouin light scattering confirms that the nonlinear response is confined to the nanoconstriction region. Lithographic control of the angle between the current and anisotropy axes tunes the excitation threshold and drives two spectral
branches from separated modes to a dominant single branch, consistent with mutual synchronization. These results establish epitaxial crystallographic anisotropy as a route to reconfigurable field-free spintronic oscillators and oscillator networks.}




\maketitle
\section*{Introduction}\label{sec1}
Spin Hall nano-oscillators (SHNOs) are nanoscale microwave sources in which spin-orbit torque converts a direct current into steady-state magnetization auto-oscillations in the gigahertz range~\cite{Demidov2012NatMater,Chen2016ProcIEEE,Dvornik2018PhysRevAppl,Zahedinejad2018ApplPhysLett,Fulara2019SciAdv,Choi2022NatCommun}. Their simple heavy-metal/ferromagnet bilayer structure, strong nonlinearity, frequency tunability, CMOS compatibility, and nanoscale footprint make them attractive for wireless communication, neuromorphic computing, and oscillator-based Ising machines~\cite{wang2019Springer,wang2021NatComput,Zhang2022micromachines,Albertsson2021ApplPhysLett,McGoldrick2022PhysRevAppl,Houshang2022PhysRevAppl,Zahedinejad2022NatMater,Zahedinejad2020NatNanotech,Demidov2014NatCommun}. In oscillator networks, SHNOs can mutually synchronize through dipolar fields or spin-wave propagation~\cite{Houshang2022PhysRevAppl,Awad2017NatPhys,Kumar2023NanoLett,Kumar2025NatPhys}, enabling collective microwave states in which oscillator phases can encode computational variables. Nanoconstriction SHNOs can be scaled to footprints of only tens of nanometres~\cite{Durrenfeld2017Nanoscale,Behera2024AdvMater}, and large lattices containing up to $10^{5}$ oscillators have recently shown coherent synchronized dynamics with quality factors exceeding $10^{6}$ and nanosecond-scale response times~\cite{behera2026NatNanotech}.

The relevant control parameter for SHNO dynamics is the total effective magnetic field, which includes contributions from external fields, magnetocrystalline anisotropy, dipolar fields, exchange, and interfacial interactions~\cite{slavin2008IEEE}. This effective-field landscape defines the equilibrium magnetization direction, sets the local resonance frequency, and determines the orientation in which spin-orbit torque can compensate damping~\cite{Fulara2019SciAdv,slonczewski1999JMMM,slavin2009IEEE}. In conventional nanoconstriction SHNOs, a large part of this landscape is supplied by an externally applied magnetic field. Such a field acts as a global bias, imposing a common magnetic axis and magnitude on all oscillators in an array~\cite{Demidov2012NatMater,Zahedinejad2018ApplPhysLett,behera2026NatNanotech}. Reliance on this global bias requires an external magnetic-field source and ties the operation of the entire array to a shared control parameter, limiting the extent to which the remanent state of each oscillator can be used as a nonvolatile degree of freedom for initialization, selection, and reconfiguration. A built-in, materials-defined contribution to the effective field would instead allow each oscillator to carry its own magnetic bias, enabling both field-free and locally programmable operation without sacrificing the simplicity of the SHNO geometry.

Field-free spin-torque auto-oscillation has already been realized in the broader family of spin-transfer-torque nano-oscillators (STNOs). In magnetic-tunnel-junction-based STNOs, zero-field microwave emission has been achieved using tilted or perpendicular polarizers, perpendicular free layers, interlayer coupling, and exchange-spring magnetic reference structures~\cite{Zhou2009njp,Zeng2013scirep,Jiang2023nanolett}. Epitaxial magnetic tunnel junctions have also been used as model spin-torque devices, where crystalline electrodes and barriers improve interface quality, spin-dependent transport, magnetic anisotropy, and microwave performance~\cite{Matsumoto2009prb}. Several strategies have been proposed to create internal effective-field landscapes for field-free SHNO operation, including engineered magnetic anisotropy~\cite{Manna2023ApplPhysLett,Gupta2024PhysRevB,manna2024PhysRevB}, exchange bias~\cite{Sravani2024IEEETrans}, demagnetizing fields~\cite{Shirokura2020JApplPhys}, and field-like torques~\cite{Arun2021JPhysCondensMat}. Among these routes, crystallographic anisotropy is particularly attractive because it is intrinsic to the ferromagnetic layer, fixed by crystal symmetry, and retained in the remanent magnetic state. In epitaxial hexagonal Co, the magnetocrystalline anisotropy associated with the crystallographic $c$-axis can generate a large in-plane effective anisotropy field when the $c$-axis is placed in the film plane~\cite{sucksmith1954RSL}. Epitaxy therefore plays a different role here from its role in conventional epitaxial STNO stacks: it defines the symmetry that supplies the internal bias field required for field-free SHNO operation.

Here, we demonstrate reconfigurable field-free auto-oscillation in nano-constriction SHNOs based on epitaxial Co/Pt bilayers grown on MgO(110). By exploiting the crystallographic anisotropy of hexagonal Co, we realize a large in-plane effective anisotropy field that replaces the external magnetic bias and enables robust auto-oscillations above 10~GHz at zero applied magnetic field. Micro-focused Brillouin light scattering shows that the auto-oscillation is confined to the nanoconstriction region, while the evolution of two spectral branches is consistent with edge-localized modes that can mutually synchronize depending on the angle between the current and anisotropy axes. The threshold current decreases as this angle approaches $90^\circ$, consistent with the angular dependence of the spin-orbit-torque efficiency. We further show that the active current polarity is selected by the remanent magnetization direction, enabling nonvolatile reconfiguration of the oscillation on/off state for a fixed current polarity. These results establish epitaxial crystallographic anisotropy as a materials-design principle for reconfigurable field-free spintronic oscillators and oscillator networks.

\section*{Results}\label{sec2}

\subsection*{Epitaxial Co/Pt provides a large in-plane crystallographic anisotropy field}\label{subsec2-1}

\begin{figure}[!b]
    \centering
	\includegraphics[width=\linewidth]{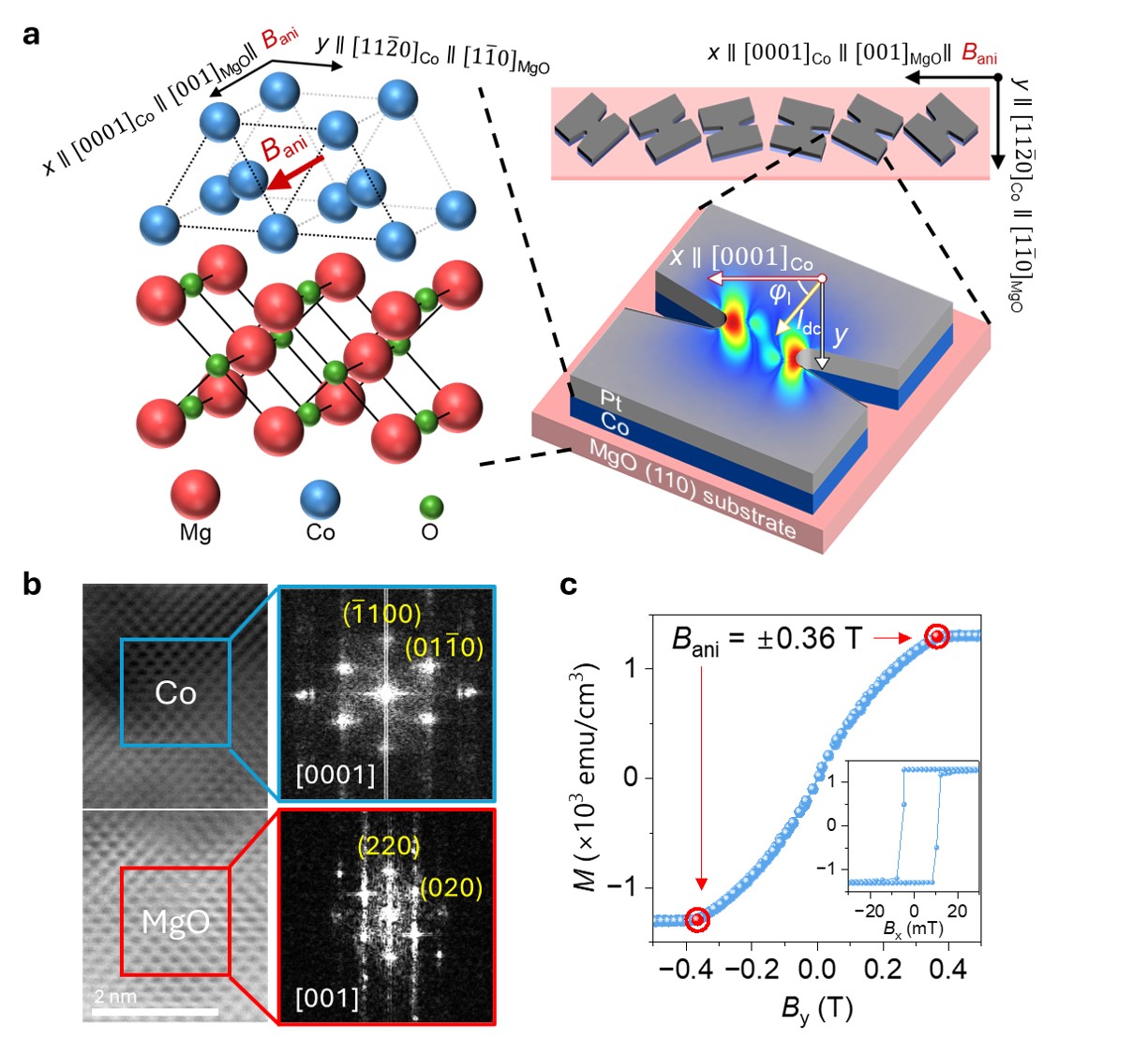}
	\caption{\textbf{Epitaxial symmetry engineering of an in-plane anisotropy field in Co/Pt.}
\textbf{a}, Atomic model of the epitaxial relation between hcp Co and MgO(110), which aligns the in-plane Co[0001] crystallographic easy axis with MgO[001] and defines the anisotropy-field direction \(B_{\mathrm{ani}}\parallel x\). Nanoconstriction SHNOs are lithographically patterned at different current angles \(\varphi_I\) relative to this fixed crystallographic axis. The enlarged device schematic shows the epitaxial Co/Pt bilayer on MgO(110); the coloured regions schematically indicate the enhanced auto-oscillation amplitude near the nanoconstriction edges.
\textbf{b}, Cross-sectional annular bright field scanning transmission electron microscopy images acquired from the Co and MgO regions, together with the corresponding fast Fourier transform patterns. The indexed reflections confirm the epitaxial orientation of the Co layer on MgO(110).
\textbf{c}, In-plane magnetic hysteresis loop measured with the field applied along the hard axis, \(y\parallel\mathrm{Co}[11\bar{2}0]\). The arrows mark the anisotropy fields \(B_{\mathrm{ani}}=\pm0.36~\mathrm{T}\). The inset shows the easy-axis loop measured along \(x\parallel\mathrm{Co}[0001]\).}
	\label{fig:F1}
\end{figure}

For spin-orbit torque to sustain auto-oscillation at zero applied field, the remanent effective-field landscape must place the magnetization in a configuration where the damping-like torque can compensate intrinsic damping. In the nanoconstriction SHNO geometry studied here, this condition is most directly achieved by an in-plane effective field with a well-defined relation to the current-induced spin polarization. We therefore use the crystallographic anisotropy of epitaxial Co to create a large in-plane effective anisotropy field, and lithographically define nanoconstrictions at selected current angles \(\varphi_I\) relative to this crystallographic anisotropy axis, as illustrated in Fig.~\ref{fig:F1}a.

The epitaxial relation between Co and MgO fixes the crystallographic origin of this effective anisotropy field. Cross-sectional scanning transmission electron microscopy and the corresponding fast Fourier transform patterns show that the Co layer grows epitaxially on MgO(110), with the Co\(\{1\bar{1}00\}\) planes aligned with the MgO\(\{220\}\) planes (Fig.~\ref{fig:F1}b). This epitaxial alignment places the hexagonal Co \(c\)-axis, Co[0001], in the film plane, defining the \(x\)-direction and the crystallographic easy axis of the ferromagnetic layer. The resulting in-plane magnetic anisotropy is confirmed by hysteresis loops measured along and perpendicular to Co[0001] (Fig.~\ref{fig:F1}c). Along the \(x\)-direction, the film exhibits an easy-axis loop with a coercive field of 8.2 mT and a finite remanent magnetization, whereas along the \(y\)-direction it shows a hard-axis loop corresponding to an anisotropy field of approximately \(0.36~\mathrm{T}\). 

\subsection*{Remanent-state-selected field-free auto-oscillation}\label{subsec2-2}

\begin{figure}[!b]
    \centering
	\includegraphics[width=\linewidth]{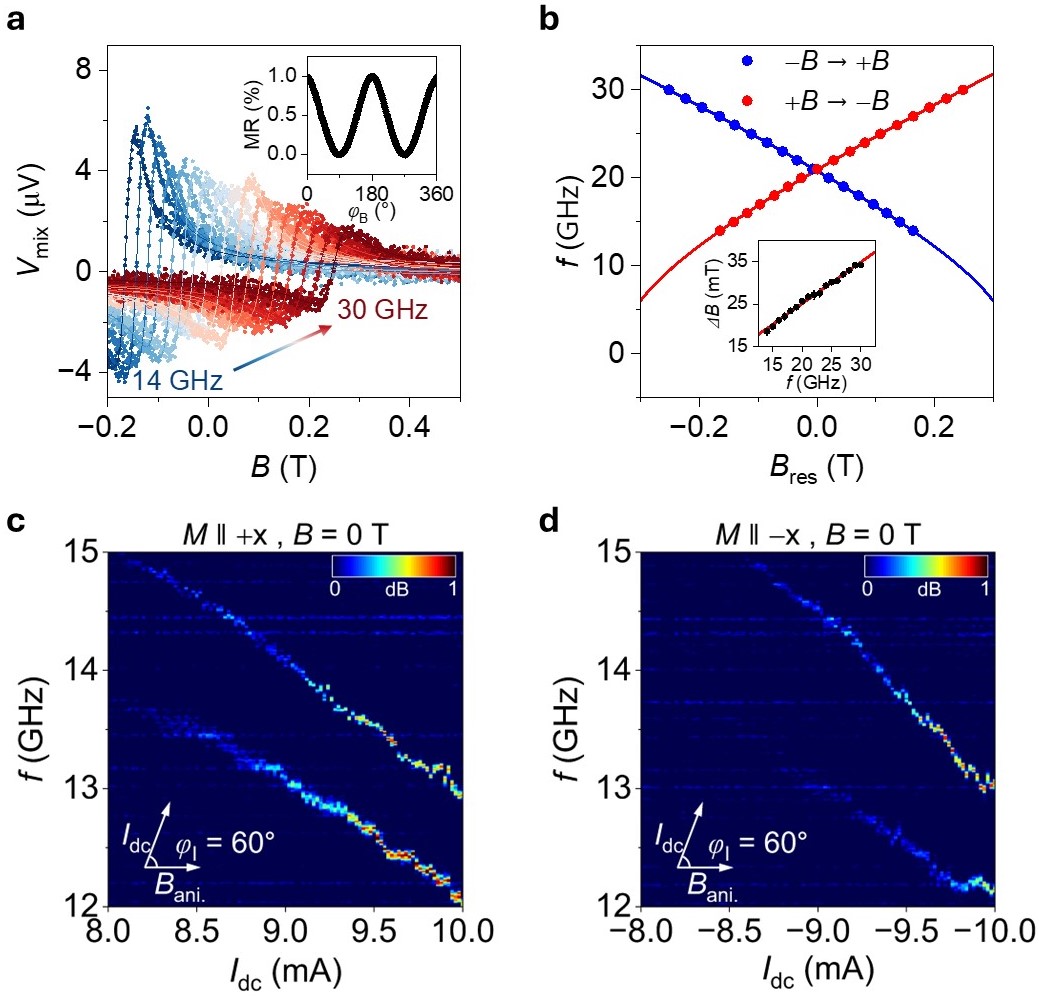}
	\caption{\textbf{Effective anisotropy field and remanent-state-selected zero-field auto-oscillation.}
\textbf{a}, Spin-torque ferromagnetic resonance (STFMR) spectra of epitaxial Co/Pt measured from 14 to 30~GHz at \(\varphi_I=60^\circ\), \(\varphi_B=0^\circ\), and \(\theta_B=0^\circ\) with a positive-to-negative magnetic-field sweep. The solid lines are fits to symmetric and antisymmetric Lorentzian line shapes. The inset shows the in-plane anisotropic magnetoresistance measured at \(B=2~\mathrm{T}\) and \(I_{\mathrm{dc}}=1~\mathrm{mA}\).
\textbf{b}, Resonance frequency \(f\) as a function of resonance field \(B_{\mathrm{res}}\) for magnetic-field sweeps from negative to positive field and from positive to negative field. The solid curves are fits to a Kittel relation including the in-plane anisotropy, yielding \(B_{\mathrm{ani}}=0.337~\mathrm{T}\). The inset shows the field linewidth \(\Delta B\) as a function of frequency; the linear fit gives an effective damping constant \(\alpha=0.0147\).
\textbf{c,d}, Current-dependent microwave power spectral density measured at \(B=0\) and \(\varphi_I=60^\circ\) after initializing the remanent magnetization along \(+x\) (\textbf{c}) and \(-x\) (\textbf{d}). Reversal of the remanent magnetization reverses the current polarity that excites field-free auto-oscillation.}
	\label{fig:F2}
\end{figure}

To establish how the crystallographic anisotropy modifies the magnetization dynamics, we performed spin-torque ferromagnetic resonance (STFMR) measurements on \(4~\mu\mathrm{m}\times8~\mu\mathrm{m}\) bars patterned from the same epitaxial Co(8~nm)/Pt(8~nm) stack. The approximately \(1\%\) AMR shown in the inset of Fig.~\ref{fig:F2}a provides electrical sensitivity to the magnetization orientation. Representative STFMR spectra measured between 14 and 30~GHz are well described by symmetric and antisymmetric Lorentzian line shapes~\cite{Liu2011PhysRevLett}. The resonance dispersions obtained for the two magnetic-field sweep directions are reproduced by a Kittel relation including the in-plane anisotropy~\cite{Manna2023ApplPhysLett}, yielding an effective in-plane anisotropy field of \(B_{\mathrm{ani}}=0.337~\mathrm{T}\). This value agrees closely with the \(0.36~\mathrm{T}\) obtained from the hard-axis hysteresis loop of the unprocessed as-deposited film stack in Fig.~\ref{fig:F1}c, showing that the large crystallographic anisotropy is largely preserved after device fabrication and patterning. The finite resonance frequency near 20~GHz at zero applied field directly shows that the crystallographic anisotropy supplies the static effective field required for high-frequency magnetization dynamics in the remanent state. The frequency dependence of the linewidth, shown in the inset of Fig.~\ref{fig:F2}b, gives an effective damping constant of \(\alpha=0.0147\).

We next tested whether this finite-frequency remanent state could be driven into auto-oscillation. Figures~\ref{fig:F2}c and ~\ref{fig:F2}d show the current-dependent microwave power spectral density measured at zero applied field after initializing the magnetization along the two crystallographic easy-axis directions. For \(M_{\mathrm{rem}}\parallel +x\), distinct microwave branches emerge above a threshold positive current and remain at frequencies above 10~GHz. Their frequencies decrease continuously with increasing current, demonstrating a pronounced negative nonlinear frequency shift. The persistence of these signals after removal of the initializing field establishes field-free auto-oscillation in the epitaxial Co/Pt nano-constriction SHNO.

Reversing the remanent magnetization to \(M_{\mathrm{rem}}\parallel -x\) reverses the active current polarity: auto-oscillation is then observed for negative current and is absent for positive current. The damping-like spin-orbit torque changes sign upon reversal of either the current or the magnetization. Consequently, reversing \(M_{\mathrm{rem}}\) reverses which current polarity reduces the effective damping and drives the system above the auto-oscillation threshold. The remanent magnetization therefore acts as a nonvolatile state variable that selects the active current polarity, providing the physical basis for the reconfigurable on/off operation demonstrated below.

The field-free spectra contain two auto-oscillation branches that lie below the corresponding ferromagnetic-resonance frequency and redshift with increasing current. These features are characteristic of nonlinear localized modes in nanoconstriction SHNOs~\cite{Dvornik2018PhysRevAppl}. The two branches are consistent with distinct localized modes associated with the two sides of the constriction, whose relative excitation and spectral evolution are examined below.

\subsection*{Angle-dependent excitation of localized auto-oscillation modes}\label{subsec2-3}

\begin{figure}[!t]
    \centering
	\includegraphics[width=12cm]{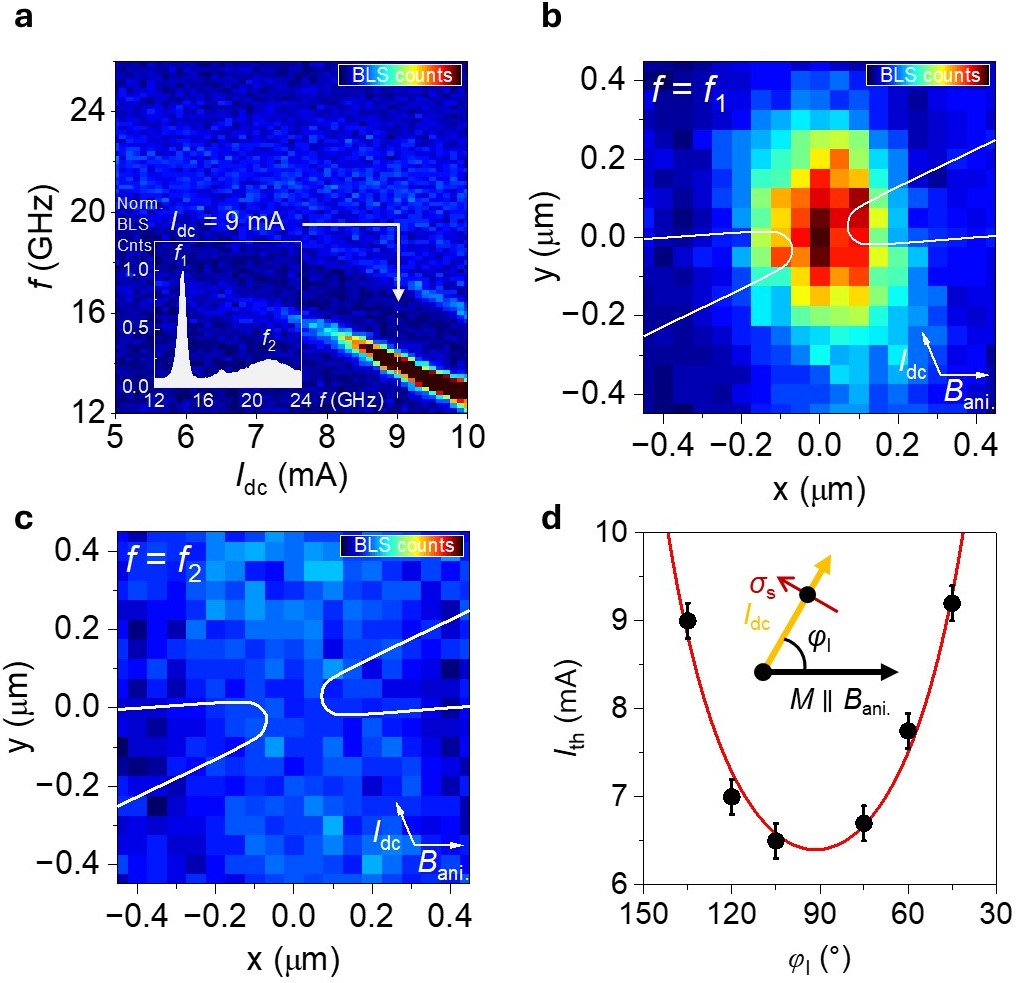}
	\caption{\textbf{Micro-focused BLS imaging of nanoconstriction-localized zero-field auto-oscillation.}
\textbf{a}, Current-dependent micro-focused Brillouin light scattering (\(\mu\)-BLS) intensity measured at \(B=0\), \(\varphi_I=105^\circ\), and \(M_{\mathrm{rem}}\parallel +x\). The inset shows the spectrum at \(I_{\mathrm{dc}}=9~\mathrm{mA}\), comprising the intense auto-oscillation mode \(f_1\) and a broader, weaker mode \(f_2\) near the ferromagnetic-resonance frequency.
\textbf{b,c}, Spatial \(\mu\)-BLS intensity maps measured at \(I_{\mathrm{dc}}=9~\mathrm{mA}\) at the frequencies \(f_1\) (\textbf{b}) and \(f_2\) (\textbf{c}) identified in the inset of \textbf{a}. The white outlines indicate the nanoconstriction geometry. The \(f_1\) intensity is concentrated within the nanoconstriction region, whereas the weaker \(f_2\) response is spatially extended.
\textbf{d}, Auto-oscillation threshold current \(I_{\mathrm{th}}\) as a function of the lithographically defined angle \(\varphi_I\) between the current and crystallographic anisotropy axes. The solid curve is a fit to \(I_{\mathrm{th}}=I_{\mathrm{th},0}/\sin\varphi_I\). The inset illustrates the relative orientations of the remanent magnetization, current, and spin Hall spin polarization. Error bars indicate the uncertainty in extracting \(I_{\mathrm{th}}\) from the current-dependent spectra.}
	\label{fig:F3}
\end{figure}

Having established field-free auto-oscillation electrically, we next examined how the orientation of the current relative to the crystallographic anisotropy axis affects the auto-oscillation dynamics. To this end, we performed micro-focused Brillouin light scattering measurements on nanoconstriction SHNOs patterned at different angles \(\varphi_I\). Figure~\ref{fig:F3}a shows the current-dependent BLS spectrum measured at \(B=0\), \(\varphi_I=105^\circ\), and \(M_{\mathrm{rem}}\parallel +x\).

The zero-field BLS spectrum contains two qualitatively different responses. A sharp and intense mode, denoted \(f_1\), emerges above \(I_{\mathrm{dc}}\approx6.5~\mathrm{mA}\) and increases rapidly in intensity with current. Its frequency decreases from approximately 18 to 12~GHz over the measured current range, consistent with the negative nonlinear frequency shift observed in the electrical spectra. A second mode, \(f_2\), appears at higher frequency with substantially lower intensity and a much broader spectral profile. Its frequency depends only weakly on current and remains close to the ferromagnetic-resonance frequency determined from STFMR, identifying it with the thermally populated magnetic response rather than the driven auto-oscillation.

Spatially resolved measurements further distinguish the two responses. Figure~\ref{fig:F3}b shows that the \(f_1\) intensity is concentrated within the nano-constriction region, where the current density and spin-orbit torque are largest. By contrast, the weaker \(f_2\) response is distributed over a substantially larger part of the device (Fig.~\ref{fig:F3}c), consistent with a spatially extended thermal mode. The spatial resolution of the present BLS measurement does not resolve the two constriction edges separately, but it confirms that the strong nonlinear \(f_1\) response originates from the nano-constriction rather than from the extended leads.

The crystallographic anisotropy also provides a direct means of controlling the excitation efficiency through the lithographically defined current angle. Figure~\ref{fig:F3}d summarizes the threshold current \(I_{\mathrm{th}}\) extracted from otherwise equivalent devices patterned at different \(\varphi_I\). The threshold decreases as the current direction approaches \(90^\circ\) relative to the anisotropy axis and is well described by
\[
I_{\mathrm{th}}=\frac{I_{\mathrm{th},0}}{\sin\varphi_I}.
\]
For a remanent magnetization aligned with the crystallographic easy axis, the projection of the spin Hall spin polarization relevant to antidamping is proportional to \(\sin\varphi_I\)~\cite{Liu2011PhysRevLett,petit2007PhysRevLett}. The measured angular dependence therefore shows that the field-free auto-oscillation threshold can be engineered directly through the orientation of the nano-constriction relative to the crystal axis.

\subsection*{Crystallographic control of edge-mode synchronization}\label{subsec2-4}

Beyond setting the auto-oscillation threshold, the orientation of the nanoconstriction relative to the crystallographic anisotropy axis also controls the interaction between the two auto-oscillation branches. To examine this effect, we measured field-dependent micro-focused BLS spectra from devices patterned at \(\varphi_I=105^\circ\), \(120^\circ\), and \(135^\circ\) (Fig.~4a-c). All measurements were performed at \(I_{\mathrm{dc}}=9.6~\mathrm{mA}\), with the magnetic field applied at \(\varphi_B=22^\circ\) and \(\theta_B=70^\circ\) and swept from positive to negative values. The magnetization initialized at positive field remains in the same remanent state as the field passes through zero and reverses only after the coercive field is exceeded. This allows the evolution of the auto-oscillation branches to be followed continuously through the zero-field state.

\begin{figure}[!t]
    \centering
	\includegraphics[width=\linewidth]{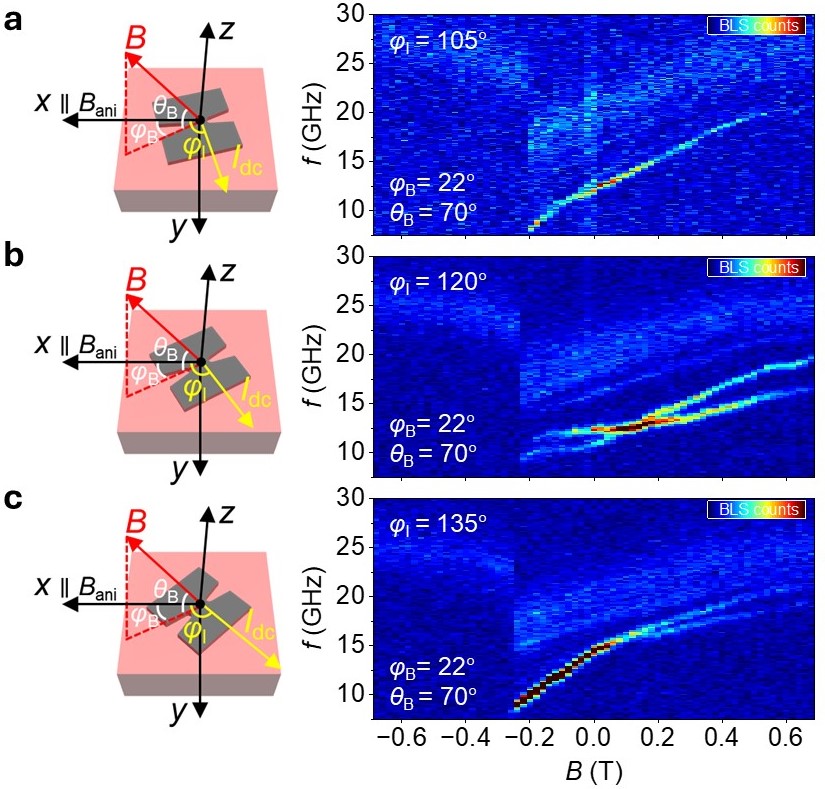}
	\caption{\textbf{Crystallographic control of edge-mode synchronization.}
\textbf{a-c}, Left, measurement geometries for nanoconstriction SHNOs patterned at current angles \(\varphi_I=105^\circ\) (\textbf{a}), \(120^\circ\) (\textbf{b}), and \(135^\circ\) (\textbf{c}) relative to the crystallographic anisotropy axis \(x\parallel B_{\mathrm{ani}}\). The applied magnetic field is oriented at \(\varphi_B=22^\circ\) in the film plane and \(\theta_B=70^\circ\) out of the film plane. Right, corresponding frequency-field \(\mu\)-BLS intensity maps measured at \(I_{\mathrm{dc}}=9.6~\mathrm{mA}\) while sweeping the field from positive to negative values. At \(\varphi_I=105^\circ\), one edge-mode branch dominates and the branches remain spectrally separated. At \(\varphi_I=120^\circ\), the two branches become comparable and merge over a finite field interval. At \(\varphi_I=135^\circ\), a single branch persists over a broad field range including \(B=0\). The progressive branch merging is consistent with angle-controlled mutual synchronization of the two edge modes.}
	\label{fig:F4}
\end{figure}

At \(\varphi_I=105^\circ\), one auto-oscillation branch dominates the spectrum, while a second, substantially weaker branch remains spectrally distinct over most of the measured field range (Fig.~\ref{fig:F4}a). At \(\varphi_I=120^\circ\), the two branches become more comparable in intensity and begin to interact more strongly at weak positive fields (Fig.~\ref{fig:F4}b). In this field range, the branches approach one another and merge into a single enhanced spectral response, accompanied by a redistribution of spectral weight between the modes. This localized spectral coalescence provides evidence of stronger mode coupling and is consistent with the onset of mutual synchronization.

At \(\varphi_I=135^\circ\), a single auto-oscillation branch persists over a substantially broader field range, including at zero applied field (Fig.~\ref{fig:F4}c). The progression from two separated branches at \(105^\circ\), through branch merging at \(120^\circ\), to a dominant single branch at \(135^\circ\) is consistent with progressively stronger mutual synchronization of the two edge-localized modes. 

\subsection*{Nonvolatile reconfiguration of field-free auto-oscillation}\label{subsec2-5}

\begin{figure}[!b]
    \centering
    \includegraphics[width=\linewidth]{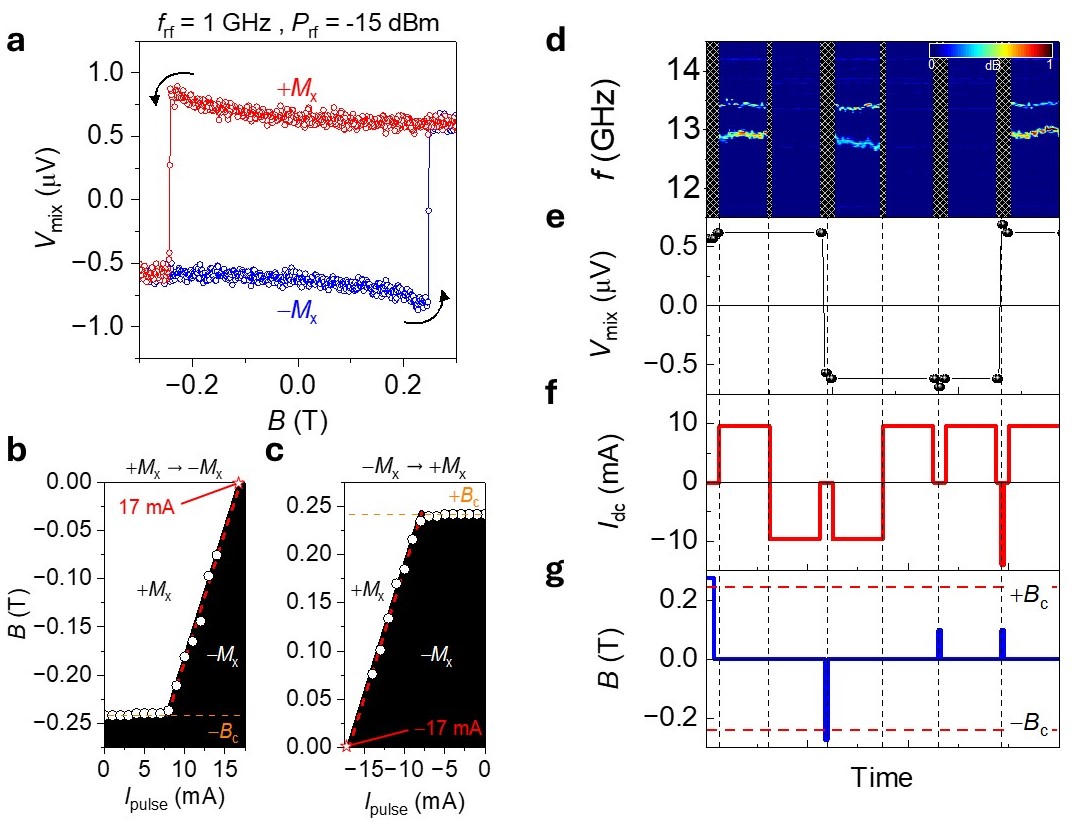}
    \caption{\textbf{Nonvolatile reconfiguration of field-free SHNO operation.}
\textbf{a}, Magnetic-field dependence of the STFMR mixing voltage \(V_{\mathrm{mix}}\), measured at \(f_{\mathrm{rf}}=1~\mathrm{GHz}\) and \(P_{\mathrm{rf}}=-15~\mathrm{dBm}\). Red and blue symbols denote field sweeps from positive to negative field and from negative to positive field, respectively. The sign of the offset voltage identifies the \(+M_x\) and \(-M_x\) remanent magnetization states.
\textbf{b,c}, Final magnetization state as a function of pulse-current amplitude \(I_{\mathrm{pulse}}\) and assisting magnetic field \(B\), starting from \(+M_x\) (\textbf{b}) and \(-M_x\) (\textbf{c}). White and black regions denote the \(+M_x\) and \(-M_x\) states, respectively, and the symbols mark the experimentally determined switching boundaries. The pulse duration is \(\tau_{\mathrm{pulse}}=70~\mu\mathrm{s}\). Dashed orange lines mark the coercive fields \(\pm B_c\). Dashed red lines are linear fits to the current-dependent portions of the switching boundaries, extrapolated to zero applied field, yielding zero-field switching currents of \(I_0^\pm=\pm17~\mathrm{mA}\).
\textbf{d-g}, Reconfiguration sequence showing the microwave power spectral density (\textbf{d}), STFMR mixing voltage (\textbf{e}), applied dc current (\textbf{f}), and applied magnetic field (\textbf{g}). The oscillator is read out at zero applied field, while current-assisted magnetization reversal is performed under a finite assisting field. Reversing the remanent magnetization reverses the active current polarity and thereby switches the field-free auto-oscillation on or off for a fixed readout current. Hatched intervals indicate the initialization or switching steps. All measurements were performed at \(\varphi_I=60^\circ\), \(\varphi_B=0^\circ\), and \(\theta_B=0^\circ\).}
    \label{fig:F5}
\end{figure}

The dependence of the active current polarity on the remanent magnetization provides a direct route to nonvolatile programming of the oscillator state. We first established an electrical readout of the two easy-axis magnetization states. Figure~\ref{fig:F5}a shows the magnetic-field dependence of the STFMR mixing voltage \(V_{\mathrm{mix}}\), measured at \(f_{\mathrm{rf}}=1~\mathrm{GHz}\). The offset in \(V_{\mathrm{mix}}\) changes sign when the magnetization reverses \cite{Yactayo2026arXiv}, allowing the \(+M_x\) and \(-M_x\) states to be distinguished electrically through the spin Seebeck effect \cite{Uchida2010ApplPhysLett} and anomalous Nernst effect \cite{nagaosa2010RevModPhys}. The hysteresis of this signal confirms that both magnetization states remain stable after the applied field is removed.

We next applied current pulses of duration \(\tau_{\mathrm{pulse}}=70~\mu\mathrm{s}\) in the presence of an assisting magnetic field. Figures~\ref{fig:F5}b and ~\ref{fig:F5}c show the final magnetization state as a function of pulse-current amplitude and magnetic field, starting from \(+M_x\) and \(-M_x\), respectively. Opposite current polarities drive the two corresponding reversal processes, and the assisting field required for switching decreases systematically as the pulse-current magnitude increases. At \(\lvert I_{\mathrm{pulse}}\rvert=14~\mathrm{mA}\), deterministic reversal requires an assisting field of approximately \(75~\mathrm{mT}\). The polarity dependence and current-induced displacement of the switching boundaries are consistent with spin-orbit torque-assisted reversal of the remanent magnetization \cite{miron2011Nature,fukami2016NatNanotech}. The pulse-current amplitudes used for magnetization reversal are substantially larger than the currents used for continuous auto-oscillation, which remain below approximately \(10~\mathrm{mA}\). We therefore restricted the switching measurements to \(\lvert I_{\mathrm{pulse}}\rvert\leq14~\mathrm{mA}\) to avoid irreversible changes in the device characteristics at higher pulse amplitudes.

The approximately linear, current-dependent portions of the switching
boundaries provide estimates of the pulse currents required for reversal at
zero applied field. Linear fits performed separately for the two switching
polarities yield extrapolated zero-field intercepts of
\(I_0^+=+17~\mathrm{mA}\) and \(I_0^-=-17~\mathrm{mA}\). These values are
empirical estimates for the fixed pulse duration of
\(70~\mu\mathrm{s}\). Micromagnetic
simulations using the experimentally determined anisotropy and device geometry
reproduce reversal of the remanent magnetization at zero applied field at
higher current density. The simulated switching occurs
between the two crystallographic easy-axis states and reverses with the
current polarity, consistent with the symmetry of the measured switching
diagrams.

Figures~\ref{fig:F5}d-g demonstrate how reversal of the remanent magnetization reconfigures the field-free oscillator. The microwave spectrum and the simultaneously measured \(V_{\mathrm{mix}}\) track the auto-oscillation and magnetization states, respectively, while the lower panels show the applied current and magnetic-field sequence. At zero applied field, \(M_{\mathrm{rem}}\parallel +x\) enables auto-oscillation for one current polarity, whereas \(M_{\mathrm{rem}}\parallel -x\) selects the opposite polarity. A programming pulse applied together with a transient assisting field reverses \(M_{\mathrm{rem}}\); after the field is removed, the new magnetization state is retained and the auto-oscillation is switched on or off for a fixed readout-current polarity. The remanent magnetization therefore acts as a nonvolatile state variable that programs the dynamical output of the oscillator.

\section*{Discussion}\label{sec3}

The crystallographic anisotropy of epitaxial Co/Pt provides a built-in in-plane effective field of approximately \(0.36~\mathrm{T}\), comparable to the external bias fields commonly used in nanoconstriction SHNOs. The close agreement between the anisotropy fields obtained from the blanket film and the patterned STFMR devices shows that this crystallographically defined effective-field landscape is largely preserved through nanofabrication. It supports finite-frequency remanent-state dynamics and auto-oscillations above \(10~\mathrm{GHz}\) at zero applied field. Reversal of the remanent magnetization reverses the current polarity for which the damping-like spin-orbit torque provides antidamping, establishing the remanent state as a nonvolatile control variable for the oscillator output. Epitaxial symmetry engineering can therefore replace a substantial part of the externally imposed magnetic-field landscape with a materials-defined contribution while introducing a non-volatile control degree of freedom that is absent from conventionally field-biased SHNOs.

The crystallographic axis also provides a lithographic means of controlling the excitation and interaction of the auto-oscillation modes. The observed \(1/\lvert\sin\varphi_I\rvert\) dependence of the threshold current shows that devices with different excitation efficiencies can be defined on the same epitaxial film simply through their orientation relative to the crystal axis. Micro-focused Brillouin light scattering further establishes that the dominant nonlinear response is confined to the nanoconstriction region. Changing \(\varphi_I\) modifies the relative excitation, detuning, and spectral coalescence of the two auto-oscillation branches. Their progression from spectrally separated modes to branch merging and a dominant single branch is consistent with progressively stronger mutual synchronization, including at zero applied field. The crystallographic orientation therefore controls both the onset of field-free oscillation and the internal multimode dynamics within an individual nanoconstriction SHNO.

The remanent magnetization also enables nonvolatile reconfiguration of the field-free oscillator. Current pulses applied with a finite assisting field reverse the magnetization and thereby switch the auto-oscillation on or off for a fixed readout-current polarity. The measured current-field switching boundaries approach zero assisting field with increasing pulse-current magnitude, and their linear extrapolation places zero-field reversal only moderately beyond the current range tolerated by the present nanoconstrictions. Micromagnetic simulations using the experimental device geometry and crystallographic anisotropy independently reproduce reversal between the two easy-axis states at zero applied field and higher current density. The programming experiments themselves therefore remain field-assisted, while the extrapolation and simulations indicate that the assisting field is imposed primarily by the current-handling limit of the present devices rather than by an intrinsic requirement of the magnetic-energy landscape. Improved nanoconstriction robustness should make fully field-free electrical programming of the remanent state accessible.

More broadly, these results establish epitaxial crystallographic anisotropy as a materials-design strategy for integrating magnetic bias, nonvolatile state selection, and synchronization control directly into spintronic oscillators. Because the excitation threshold and mode interaction can be adjusted lithographically while the remanent magnetization provides a programmable local state, the approach offers a route towards oscillator arrays in which individual elements can be initialized, selected, and reconfigured without a continuously applied global magnetic field. Combining such local programmability with dipolar or spin-wave-mediated coupling between oscillators could enable field-free networks whose active elements and collective dynamical states are defined jointly by crystal symmetry, device orientation, and nonvolatile magnetic configuration.

\section*{Methods}\label{sec4}
\subsection*{Material growth}\label{subsec41}
Epitaxial Co (8 nm)/Pt (8 nm)/$\mathrm{AlO_x}$ (2 nm) layers were deposited on MgO (110) single-crystalline substrates using an ultrahigh-vacuum magnetron sputtering system with a base pressure of $1\times10^{-6}$ Pa. MgO (110) substrates were annealed at $850^\circ\mathrm{C}$ for 2 hours under an oxygen pressure of 8.0 Pa to reconstruct atomically flat surfaces. A 4 nm Co layer was deposited at a substrate temperature of $250^\circ\mathrm{C}$, whereas the other 4 nm of the Co layer was deposited at room temperature. The Co layer was deposited using a sputtering power of 30 W with a working pressure of 0.4 Pa. The Pt layer was deposited at 5 W with a working pressure of 0.4 Pa at room temperature. All the metallic layers were grown using DC sputtering, while the amorphous $\mathrm{AlO_x}$ capping layer was grown by RF sputtering (40 W) at 0.33 Pa at room temperature.

\subsection*{Materials characterization}\label{subsec42}
Scanning transmission electron microscopy (STEM) cross-sections were prepared using a 30 kV $\mathrm{Xe^+}$ plasma-focused ion beam using the Thermo Fisher Scientific Helios 5 Hydra CX DualBeam Plasma FIB/SEM. Final thinning and cleaning steps were performed using 8 kV and 5 kV $\mathrm{Xe^+}$ ion beam as well. The samples were additionally polished using a Fischione Model 1040 Nano Mill at 1- 0.5 kV to remove residual amorphization and damage, then Ar plasma-cleaned using a South Bay Technology PC-2000 Plasma Cleaner at 20- 40 W RF power for 15 s at 20 Pa. Scanning transmission electron microscopy (STEM) was performed at 200 kV using an aberration-corrected JEOL JEM-ARM200CF S/TEM, with a 27 mrad convergence angle and a 23 mrad annular bright-field (ABF) collection angle. Data processing (denoising, fast Fourier transform, Fourier filtering) was performed using the Gatan Microscopy Suite (GMS) software.

Magnetic hysteresis loops were measured at 300 K using a vibrating-sample magnetometer (VSM) option of a Physical Property Measurement System (PPMS DynaCool, Quantum Design). The in-plane magnetic field was swept along the MgO $[1\bar{1}0]$ and MgO $[001]$  directions to maximum fields of ± 1 T  and ± 0.1 T, respectively. The linear diamagnetic background from the substrate was subtracted.

\subsection*{Device fabrication}\label{subsec43}
Spin Hall nano-oscillator (SHNO) and spin-torque ferromagnetic resonance (STFMR) bar-shaped devices were fabricated using electron beam lithography (EBL) with an overall dimension of $4~\mu$m $\times$ $8~\mu$m. In particular, SHNO devices incorporated a nano-constriction structure at the center, designed with a radius of 50 nm and an opening angle of $22^\circ$, resulting in a constriction width of 150 nm. The EBL process was carried out by first coating the material stack with a negative resist (MaN 2401), followed by exposure using a Raith EBPG 5200 system. Subsequently, Ar-ion beam etching was performed using an Oxford Ionfab 300 Plus etcher. Ground-signal-ground (GSG) coplanar waveguides (CPWs) were then defined through an optical lithography lift-off process on a Au (250 nm) deposition.

\subsection*{Electrical measurement}\label{subsec44}
For spin torque ferromagnetic resonance (STFMR) measurement, the ground-signal-ground coplanar waveguide (GSG-CPW) of device, signal generator (Rohde \& Schwarz SMB100A), and lock-in amplifier (SR830) together with a source meter (Keithley 6221) were connected to the DC+RF port, RF port, and DC port of the bias-T, respectively. Microwave signals with powers of 4 dBm or $-15$ dBm generated by the signal generator with amplitude modulation at a reference frequency of 213.5 Hz were applied to the sample. The mixing voltage locked to the reference frequency was subsequently detected using the lock-in amplifier. The external magnetic field was applied to the x-direction (easy axis of the Co layer) through a custom-built probe station. For magnetization switching experiments, current pulses with a duration of $70~\mu$s were applied using the source meter during STFMR measurement.

The power spectral density of magnetization auto-oscillation was measured using a custom-built probe station. The direction of the external magnetic field was controlled by rotation of the device. The ground-signal-ground coplanar waveguide (GSG-CPW) of the SHNO was connected to the DC+RF port of the bias-T. DC current from the source meter (Keithley 6221) was fed to the device through the DC port of the bias-T. The auto-oscillation signal generated from the SHNO was transmitted through a low-noise amplifier (Gain = 36 dB, noise figure = 0.78 dB) connected to the RF port of the bias-T and subsequently recorded using a Rohde \& Schwarz FSV40 spectrum analyzer.

\subsection*{Microfocused-Brillouin light scattering ($\mu$-BLS) measurement}\label{subsec46}
A monochromatic continuous-wave laser with wavelength
\(\lambda=532~\mathrm{nm}\) was focused onto the samples by a $\times100$ microscope objective (MO) with a large numerical aperture (NA = 0.75), down to a 300 nm diffraction-limited spot diameter. The external magnetic field $B$ was applied at an in-plane angle ($\varphi_\mathrm{B}$) of $22^\circ$ and an out-of-plane angle ($\theta_\mathrm{B}$) of $70^\circ$. 
The inelastically scattered light from the sample was collected by the same MO and analyzed using a Sandercock-type six-pass tandem Fabry-Perot interferometer (TFP-1, JRS Scientific Instruments)~\cite{demokritov2007micro}. A stabilization software based on an active feedback algorithm (THATec Innovation) was employed to achieve long-term spatial stability during the $\mu$-BLS measurements. The experimentally obtained BLS intensity is proportional to the square of the amplitude of the dynamic magnetization.

\subsection*{Micromagnetic simulations}\label{subsec47}
The micromagnetic simulations of the ferromagnetic layer were performed using the GPU-accelerated MuMax3 software~\cite{vansteenkiste2014design}. The simulated geometry consisted of a $400 \times 600 \times 8$ nm ferromagnetic layer with a 150 nm-wide constriction, discretized into $400 \times 600 \times 1$ cells. The material parameters used in the simulations were obtained from sample measurements, including a saturation magnetization of $M_s = 1.302 \times 10^6$ A/m, an exchange stiffness coefficient of $A_\mathrm{ex}=1\times10^{-11} \mathrm{J/m}$, an in-plane anisotropy constant of $K_\mathrm{u} = 0.22 \times 10^6$ J/m$^3$, a Gilbert damping constant of $\alpha = 0.0147$, and a spin-Hall angle of $\theta_{\textbf{SH}} = 0.13$ (taken from literature~\cite{Choi2017PhysRevB}).
The out-of-plane spin-polarized currents were calculated using a 2D steady-state conduction problem that accounts for the device geometry, layer thicknesses, and the electrical properties of Co and Pt. The parameters used for the current-density calculation were obtained from sample measurements: the Co resistivity $\rho_{Co} = 23.68 ~\mu\Omega\, \mathrm{cm}$, Co thickness $t_{Co} = 8$ nm, Pt resistivity $\rho_{Pt} = 83  ~\mu\Omega\,\mathrm{cm}$, and Pt thickness $t_{Pt} = 8$ nm. 

\backmatter

\bmhead{Acknowledgements}
This work was partially supported by the Swedish Research Council (VR Grant No. 2024-01943) and the Knut and Alice Wallenberg Foundation (Grant No. 2023.0285). The work at Northwestern University was in part supported by the Center for Energy-Efficient Magnonics (CEEMag), an Energy Frontier Research Center funded by the U.S. Department of Energy (DOE), Office of Science, Basic Energy Sciences (BES), under award number DE-AC02-76SF00515 (material growth and characterization) and by the U.S. National Science Foundation (NSF) under award number 2203242 (magnetic characterization and data analysis). This work also made use of the EPIC facility (RRID: SCR\_026361) at Northwestern University’s NUANCE Center, which is supported by the International Institute for Nanotechnology (IIN) and the Northwestern University Materials Research Science and Engineering Center (MRSEC) program (NSF DMR-2308691).

\bmhead{Author contribution}
J.-G.C. initiated the idea. J.-G.C., A.K.C., and J.K. designed the study. J.K. grew, optimized and characterized the epitaxial thin film stacks with P.G.L, M.C.H, V.P.D and P.K.A. J.-G.C. fabricated devices and performed all electrical measurements. A.K.C. performed all microfocused BLS measurements. Both J.-G.C. and A.K.C. analyzed the data with A.K., A.A.A., and J.Å. V.V. performed the micromagnetic simulation in discussion with R.K., J.-G.C., and A.K.C. All co-authors contributed to the manuscript, the discussion, and the analysis of the results. J.Å. and P.K.A. managed and supervised the project.

\bibliography{sn-bibliography}

\end{document}